\documentclass[11pt]{article}
\usepackage[latin1]{inputenc}
\usepackage[english]{babel}
\usepackage[namelimits]{amsmath}
\usepackage{authblk}
\usepackage{amssymb}
\usepackage{amsmath}
\usepackage{amsthm}
\usepackage{graphicx}

{\tiny}

\begin{document}
\title{{\bf Physically viable rotating mass solutions surrounding Kerr black holes}}
\author[1,2,3]{S. Viaggiu\thanks{s.viaggiu@unimarconi.it and viaggiu@axp.mat.uniroma2.it}}
\affil[1]{Dipartimento di Scienze Ingegneristiche, Universit\'a degli Studi Guglielmo Marconi, Via Plinio 44, I-00193 Roma, Italy.}
\affil[2]{INFN, Sezione di Roma 3, I-00146 Rome Italy.}
\affil[3]{Dipartimento di Matematica, Universit\`a di Roma ``Tor Vergata'', Via della Ricerca Scientifica, 1, I-00133 Roma, Italy.}
\date{\today}\maketitle

\begin{abstract} 
There exists in literature an increasing interest in the study of mass distributions surrounding black holes as describing dark matter halo in spiral galaxies. Motivated by this interest, we study a very recent new class of rotating solutions that are suitable to build anisotropic matter sources surrounding rotating black holes. Contrary to the mainstream approach, instead of use the so called regular black holes as central objects, we perform a smooth matching between the aforementioned anisotropic matter and a central vacuum Kerr black hole. In this framework, we study in full generality energy conditions near the matching surface. As a result, we found that, after imposing the vanishing of the energy density
$E$ at the matching surface, if weak and dominant energy conditions (WEC,DEC) are satisfied, then unavoidable strong energy conditions is violated, i.e. near the event horizon only matter with dark energy-like features is allowed. As an application, we present two solutions everywhere satisfying 
DEC. The first one is asymptotically flat and equipped with a non vanishing electric charge, while the second solution presented is equipped with a non-vanishing 
energy flow around the symmetry axis and it is not asymptotically flat.

\end{abstract}
 Keywords; galaxy models, rotating fluids, anisotropic fluids, Kerr black holes,
energy conditions

\section{Introduction}

Many celestial bodies in the universe are rotating, It is thus of great astrophysical interest to find
exact metrics describing rotating, axially symmetric, isolated bodies. 
In the literature, many methods have been developed  
(see for example \cite{1}-\cite{27} and references therein) to build physically viable rotating metrics for rotating astrophysical objects. To this purpose, since of the pioneering work in \cite{28}, anisotropic sources have attracted a lot of interest.
In particular, in \cite{29} it has been shown
that, at least in the limit of slow rotation, the equation of hydrostatic equilibrium for a rotating source is provided
by anisotropic fluids.
In \cite{30} it has been shown that, both for spherical and axially symmetric 
astrophysical sources, stellar evolution processes, also for initial isotropic conditions, 
will always tend to form anisotropies in the fluid composing the source.\\ 
The recent achievement of supermassive black hole (SMBH) at the centre of M87 \cite{31} and
Sgr A* (our galaxy) \cite{32} by Event Horizon Telescope Collaboration increased the interest and the 
necessity to obtain theoretical models mimicking a central SMBH surrounded by a dark matter halo, with particular attention payed to the shadow images of SMBH and 
orbital motion of the well known S2 star near the event horizon of SMBH at Sgr A* (see for example \cite{33}-\cite{43} and references therein). In mainstream approaches, the central SMBH is often depicted in terms of static solutions (see for example \cite{43}) or regular rotating black holes 
(see for example \cite{34}, \cite{44}-\cite{50} and references therein). Regular rotating black holes are obtained by means of Gurses-Gursey algorithm. The regular solutions generated with the aforementioned 
algorithm generally suffer for a violation of WEC, typically in some region within the event horizon
\cite{44}. It is widely accepted that a violation of WEC is possible only at a quantum level: no macroscopic matter is known to violate WEC. Since regular non-vacuum black hole solutions are 
justified by the necessity to solve classically singularity, the presence of matter with negative energy 
density can be considered a weakness of these models. In a more conservative approach, we may think to retain the vacuum Kerr solutions for the central SMBH and perform a matching at some radius $r$ with
an exterior non-vacuum solution. In fact, it is expected that some quantum gravity phenomenon, for example due to possible non commutative effects at Planckian scales \cite{51}-\cite{54}, could solve singularity. If this is the case, at a macroscopic level we can certainly retain the classical Kerr solution to depict the central black hole. To this purpose, we can advantage of a new class of rotating anisotropic fluids in
\cite{55} containing the Gurses-Gursey metrics as a subcase. By means of the new metric in \cite{55},
we can obtain and study new rotating anisotropic fluids smoothly matched to the Kerr one on a spheroidal surface near the event horizon. The main purpose of this paper is not to exactly describe rotation 
galaxy curves, but rather to give a general study of the consequences of usual and reasonable matching and boundary conditions on energy conditions. Energy conditions are an important tool to understand the possible nature of dark matter halo surrounding  SMBH of spiral galaxies.\\
In section 2 we present mathematical preliminaries. In section 3 we study energy conditions for solutions
in \cite{55} with a non-vanishing energy flow around the symmetry axis, while in section 4 we analyse solutions belonging to the Gurses-Gursey class. In sections 5 and 6 we present and study 
explicit solutions. Finally, section 7 is devoted to conclusions and final remarks.

\section{Mathematical preliminaries: metric, boundary conditions and matter content}

The starting point for this paper is the new metric 
in \cite{55} describing rotating anisotropic fluids.
In the Boyer-Lindquist coordinates $(t,r,x,\phi)$ with 
$\rho=\sqrt{r^2+a^2}\sqrt{1-x^2}, z=rx$ with $x=\cos\theta$, the aforementioned new metric 
in \cite{55} is given by:
\begin{eqnarray}
& & ds^2=\label{1}\\
& & -\left(1-\frac{2m(r)f(r)}{\Sigma}\right)dt^2+\Sigma\left(\frac{dx^2}{1-x^2}
+\frac{dr^2}{\Delta}\right)-\frac{4m(r)f(r)a(1-x^2)}{\Sigma}dt d\phi+\nonumber\\
& &+\left({f^2(r)}+a^2+\frac{2m(r)f(r)a^2}{\Sigma}(1-x^2)\right)(1-x^2)d\phi^2, \nonumber\\
& & \Sigma=f^2(r)+a^2x^2,\;\;\Delta=f^2(r)-2m(r)f(r)+a^2. \label{2}
\end{eqnarray}
Solution (\ref{1}) reduces to the vacuum Kerr solution for $f(r)=r$ and $m(r)=m$, where $m$ is the ADM mass of the black hole. 
Moreover, with the transformation $m(r)\rightarrow m(r)-Q^2/(2f(r))$, with $Q$ an electric charge, metric
(\ref{1}) can also include an electric field \cite{55}. 
In order to depict a central Kerr black hole surrounded by a matter field, we must 
perform a smooth matching between the Kerr metric and (\ref{1}) on a given spheroidal surface located 
at $r=R$. For the continuity of the first and the second fundamental form on $r=R$ we must have:
\begin{equation}
f(R)=R,\;f_{,r}(R)=1,\;m(R)=m,\;m_{,r}(R)=0.
\label{3}
\end{equation}
Moreover, since outer event horizon is located at $R_H=m+\sqrt{m^2-a^2}$, we must have $R>R_H$.\\
As commented in \cite{55}, on general grounds it is required for $f(r)$ to  be a monotonically increasing
function. Moreover, in order to preserve the physical meaning of $m(r)$ as a mass function, we must 
choose also for $m(r)$ a monotonically increasing function.\\
After introducing a suitable  orthonormal tetrad  
$\{V_{(t)}^{\mu}, S_{(r)}^{\mu}, L_{(x)}^{\mu}, W_{(\phi)}^{\mu} \}$ with
\begin{equation}
g_{\mu\nu}=-V_{(t)\mu}V_{(t)\nu}+S_{(r)\mu}S_{(r)\nu}+L_{(x)\mu}L_{(x)\nu}+
W_{(\phi)\mu}W_{(\phi)\nu},
\label{4}
\end{equation}
where
\begin{eqnarray}
& & V_{(t)}^{\mu}=
\left[\frac{{f(r)}^2+a^2}{\sqrt{\Sigma\Delta}}, 0, 0, \frac{a}{\sqrt{\Sigma\Delta}}\right],
\label{4a}\\
& & S_{(r)}^{\mu}=
\left[0, \sqrt{\frac{\Delta}{\Sigma}}, 0, 0\right],\nonumber\\
& & L_{(x)}^{\mu}=
\left[0, 0, \sqrt{\frac{1-x^2}{\Sigma}}, 0\right],\nonumber\\
& & W_{(\phi)}^{\mu}=
\left[\frac{a(1-x^2)}{\sqrt{\Sigma(1-x^2)}}, 0, 0, \frac{1}{\sqrt{\Sigma(1-x^2)}}\right].
\nonumber
\end{eqnarray}
The energy-momentum tensor $T_{\mu\nu}$ generated by (\ref{1}) with (\ref{4a}) is given by
\begin{eqnarray}
& & T_{\mu\nu}=EV_{(t)\mu}V_{(t)\nu}+P_rS_{(r)\mu}S_{(r)\nu}+P_xL_{(x)\mu}L_{(x)\nu}+
P_{\phi}W_{(\phi)\mu}W_{(\phi)\nu}+\nonumber\\
& &+ K\left(V_{(t)\mu}W_{(\phi)\nu}+V_{(t)\nu}W_{(\phi)\mu}\right),
\label{5}
\end{eqnarray}
where $E$ is the energy density, $\{P_r, P_x, P_{\phi}\}$ principal stresses and 
with $K$ denoting  a non-vanishing energy flow around the symmetry axis. We also have
\begin{eqnarray}
& & E+P_r=-2\frac{\Delta\left[f_{,r,r}f\;\Sigma+
a^2 x^2(1-f_{,r}^2)\right]}
{\Sigma^3},\label{6}\\
& & P_{x}-P_{\phi}=\frac{2a^2f^2(1-x^2)(f_{,r}^2-1)}{\Sigma^3}, \label{7}\\
& & K=\frac{a\sqrt{\Delta(1-x^2)}\left[f f_{,r,r}-1+f_{,r}^2\right]}{\Sigma^2}.\label{8}
\end{eqnarray}
Energy conditions can be obtained in terms of eigenvalues of
$T_{\mu\nu}$ and consequently the equation to solve is $|T_{\mu\nu}-\lambda g_{\mu\nu}|=0$.
Hence, for the WEC we have:
\begin{equation}
-\lambda_{t}\geq 0,\;\;\;-\lambda_{t}+\lambda_{i}\geq 0,
\label{9}
\end{equation}
for the SEC:
\begin{equation}
-\lambda_{t}+\sum_{i}\lambda_{i}\geq 0,\;\;\;-\lambda_{t}+\lambda_{i}\geq 0,
\label{10}
\end{equation}
while for DEC:
\begin{equation}
-\lambda_{t}\geq 0,\;\;\;\lambda_{t}\leq \lambda_{i}\leq -\lambda_{t},
\label{11}
\end{equation}
where
\begin{eqnarray}
& & \lambda_{t}=
\frac{1}{2}\left(P_{\phi}-E-\sqrt{{(E+P_{\phi})}^2-4K^2}\right),\label{12}\\
& & \lambda_{\phi}=
\frac{1}{2}\left(P_{\phi}-E+\sqrt{{(E+P_{\phi})}^2-4K^2}\right),\label{13}\\
& & \lambda_{r}=P_{r}, \label{14}\\
& & \lambda_{x}=P_x. \label{15}
\end{eqnarray}
From a physical point of view, WEC (\ref{9}) does imply that energy density is always positive for any
timelike physical observer. SEC (\ref{10}) means that pressures can be negative but not too negative
to violate inequality (\ref{10}). 
Moreover, DEC does imply that speed velocity of a given fluid cannot
exceed the speed of the light. A violation of DEC does imply a violation of causality: no violation of causality has been observed until now in our universe.  
Furthermore, it should be stressed that no classical matter has been 
until now observed
in the universe violating WEC and DEC. In fact, it is believed that WEC can be violated only at a quantum
level (Casimir effect) and consequently no stable macroscopic matter can exist violating WEC.
Conversely, the current standard cosmological model, dubbed LCDM, predicts the existence of dark energy
in terms of a small cosmological constant $\Lambda$ 
and thus violation of SEC does not compromise the 
physical viability of a model.
Dark energy satisfies WEC but violates SEC.\\
Consequently, it is expected that also matter content near a SMBH fulfills WEC and DEC but could violate
SEC.\\
In practice, it is assumed that for physical viability at a classical level, where General Relativity
certainly does apply, at least WEC cannot be violated, also for matter living very near 
the event horizon where huge physical effects of SMBH come into action.
As a result, the aim of our study is to obtain mathematical conditions assuring the non violation of, at least, WEC (and also DEC). In the case that WEC is violated, we indicate the way we can 'cure' such a 
violation by adding addictional fields. The goal is thus to find physically viable models, satisfying   at least WEC and DEC, that can be used as toy models to study the features of matter-radiation near a SMBH, in order, for example, to explore the main features of dark matter surrounding a SMBH.\\
Hence, we analyze the energy conditions under general conditions near the event horizon of  
a SMBH. As stated at the introduction, anisotropic matter seems to be the one appropriate to describe 
rotating matter. This study can allow to obtain simple but viable models to study, for example, 
dark matter halo near SMBH: to this purpose,
we must consider further conditions for $E$ on the boundary surface
$r=R$. In effect, as shown in \cite{41b}, in order to consider a realistic situation where a black hole is surrounded by a possible dark matter halo, by performing a Newtonian study, we must have 
$E(R=4R_s=8m)=0$. This result has been amended in \cite{42} by using a Schwarzschild metric in a general relativistic context, with $E(R=2R_s)=0$. The result in \cite{42} is physically reasonable since
$R=2R_s$ is the radius $R_{ms}$
of the unstable circular orbit in the static case for a marginally bound particle.
In the rotating case, formula for $R=2R_s=R_{ms}$ should be corrected to:
\begin{equation}
R_{ms}=m{\left(1+\sqrt{1 \pm a/m}\right)}^2,
\label{15a}
\end{equation} 
where in (\ref{15a}) the plus sign refers to retrogae motion for matter field, while minus sign refers to 
prograde matter. For $a=0$ we regain the Schwarzschild value $R=2R_s$.\\
However, as we will see in the next section, the exacy location of $R_{ms}$ does not play a fundamental role for the purposes of this article.\\
As a final consideration of this section, it should be also stressed that, in order to study SMBH,
we do not consider models with regular BH. We instead use usual Ker metric to depict SMBH, this is because it is the most simple and natural choice we can made. However, there exists another important reason to use the aforementioned modelling. In fact, our models allow to study possible physical consequence of using matching conditions at $r=R$. Hence, the role of matching conditions can be studied
in order to outline possible differences with respect to usual modelling using regular BH.
To this purpose, see for example discussion below equation (\ref{29}).
Stated in other words, the use of Kerr BH does impose matching conditions at $r=R$ and this leads to models that can be physically different compared with the ones obtained with regular BH (as an example shadow images can be different and this could discriminate between a Kerr or a regular BH at the center of our galaxy).\\ 
We are now in the position to start the study of energy conditions, at least near $r=R$.

\section{Energy conditions: the case $f(r)\neq r$}

In our approach, metric (\ref{1}) must be smoothly matched to vacuum Kerr black hole solution on 
$r=R$ by imposing conditions (\ref{3}), together with the physical requirement for a dark matter halo
\cite{41b,42} $E(R=R_{mb})=0$. It is thus necessary to study the consequences of these boundary conditions for the fulfillment of energy conditions. As a first consideration, note that in
(\ref{12})-(\ref{15}) terms linear in $a$ are absent.\footnote{The function $K$ ia proportional to $a$ but appears quadratically in $\lambda_{t}$.} Hence, in the slowly rotating limit, obtained from
(\ref{1})-(\ref{2}) by dropping quadratic terms in the parameter $a$, energy conditions look like the static case obtained by setting $a=0$. As a result, for energy conditions we can study both static and slowly rotating limit by setting $a=0$.\\ 
The first step is to study energy conditions near $r=R>2m$. There, we must 
thus specify the behavior of $f(r)$ and
$m(r)$. 
First of all, we may suppose that $f(r)$ and $m(r)$ admit $n$ derivatives in $r=R$. In
such a case, thanks to Peano's theorem, both $f(r), m(r)$ can be approximated by means of 
Taylor polynomials:
\begin{eqnarray}
& & f_n(r)=\sum_{k=0}^{n}\frac{f^{(k)}(R)}{k!}{(r-R)}^k,\;
 m_n(r)=\sum_{k=0}^{n}\frac{m^{(k)}(R)}{k!}{(r-R)}^k,\label{16}\\
& & f(r)=f_n(r)+o({(r-R)}^n),\;m(r)=m_n(r)+o({(r-R))}^n.\label{17} 
\end{eqnarray}
According to (\ref{16})-(\ref{17}) we can write:
\begin{eqnarray}
& & f(r)=r+B{(r-R)}^2+C{(r-R)}^3+D{(r-R)}^4+o({(r-R)}^4),\label{18}\\
& & m(r)=m+G{(r-R)}^2+H{(r-R)}^3+L{(r-R)}^4+o({(r-R)}^4),\label{19}
\end{eqnarray}
where $\{B,C,D,G,H,L...\}\in\Re$. Series expansions (\ref{18}) and (\ref{19}) obviously satisfy
boundary conditions (\ref{3}). Moreover, the general condition $m_{,r}(r)>0$ requires, with
$G\neq 0$, that $G>0$  near $r=R$. 

\subsection*{Static and slowly rotating limit}

As stated above, for the slowly rotating limit we can set $a=0$. In that limit we have
$K=0$ and $P_{\phi}=P_x$
As a first step, we analyse the limit for $r\rightarrow R$ for $E$ and $E+P_{\phi}$ with 
(\ref{18}) and (\ref{19}):
\begin{equation}
E(R)=-\frac{4B}{R^2}(R-2m),\;\;E(R)+P_{\phi}(R)=
-\frac{2}{R^2}\left[RG+B(R-3m)\right]. 
\label{20}
\end{equation}
From (\ref{20}) we have that condition $E(R)=0$ \cite{41b,42} for a dark matter halo near the outer event horizon requires $B=0$. If this is the case, we must have $G<0$, but for a monotonically increasing ecpression for $m(r)$ we must have $G>0$. Hence, we also have $G=0$. As we see later in this section this is a general fact: in order to satisfy matching conditions (\ref{3}) and $E(R)=0$ the degree 
of the principal part of $f(r)$ given by (\ref{18}) cannot be greater than the one for $m(r)$
given by (\ref{19}). We also stress again that, if we take $R=R_{ms}$, where $R_{ms}$ is 
the radius 
of the unstable circular orbit for a marginally bound particle, it is thus expected that for 
$r\leq R=R_{ms}$ no particles can stably survive. As a result, we can take for $r\leq R$ the vacuum 
Kerr solution with obviously $R>2m$. 
Moreover, note that, since $P_{\phi}(R)=-2G/R$, by setting $B=0$ and $G<0$, 
since $\lim_{r\rightarrow R} P_{\phi}(r)/E(r)$ is diverging, 
DEC is certainly violated at and near $r=R$.\\
With respect to (\ref{18}) and (\ref{19}) with $B=G=0$, for
the relevant quantities for WEC and SEC we have:
\begin{eqnarray}
& & E=-\frac{12 C}{R^2}(R-2m)(r-R)-\label{21}\\
& & -\frac{6}{R^3}
\left[-RH-C(R-8m)+4DR(R-2m)\right]{(r-R)}^2+o(1),\nonumber\\
& & E+P_{\phi}=\frac{6}{R^2}\left[-C(R-3m)-RH\right]{(r-R)}+\label{22}\\
& & +\frac{12}{R^3}\left[-3mC+RH-RD(R-3m)-LR^2\right]{(r-R)}^2+o(1),\nonumber\\
& & E+P_r=-\frac{12C}{R^2}(R-2m)(r-R)+\label{23}\\
& & +\frac{12}{R^3}\left[C(R-4m)-2RD(R-2m)\right]{(r-R)}^2+o(1),\nonumber\\
& & E+P_r+2P_{\phi}=\frac{12}{R^2}(mC-RH)(r-R)+\label{24}\\
& &+\frac{12}{R^3}
\left[RH-2mC+2RmD-2R^2L\right]{(r-R)}^2+o(1),\nonumber
\end{eqnarray}
with $o(1)=o({(r-R)}^2)$. From (\ref{21}), for $C\neq 0$, we must have $C<0$. 
From (\ref{23}) we have that for $r\rightarrow R$, $E+P_r\geq 0$ is satisfied. 
From (\ref{24}) we deduce that $H<0$, but for the physical request $m_{,r}>0$ we must
have $H>0$. Consequently in the case $C\neq 0$ SEC is violated at least in a neighborhood of
$R$. For $C=0$, from (\ref{22}) we have $H<0$ and WEC is violated. Moreover, from 
(\ref{22}) for $C<0$ and $R>3m$ and $H>0$, condition $E+P_{\phi}$ can be certainly satisfied with 
$|C|\geq RH/(R-3m)$.
For $H=0$ and $C<0$ from (\ref{24}) we have a violation of SEC. Finally for 
$C=H=0$, from (\ref{21}) we must have $D<0$, but  SEC form (\ref{24}) requires that
$L<0$ that once again violates the condition $m_{,r}\geq 0$.\\
From this preliminary study we can deduce that 
when WEC is satisfied sufficiently near $r=R$, unavoidably SEC is violated. 
Concerning DEC near $R$, we have that $\lim_{r\rightarrow R} p_r/E=0$. Furthermore,  
with (\ref{18})-(\ref{19}) we have
$\lim_{r\rightarrow R}|P_{\phi}|/E=|(-HR+C(R-m))|/(2|C|(R-2m)<1$. The last inequality can be easily satisfied for suitable values of $C,H$. As an example, for $H=0$ we must have $R>3m$ that is certainly satisfied thanks to our choice for $R$.\\
To generilise the reasonings above, we can consider the following expressions for $f(r)$ and $m(r)$ for 
$r\rightarrow R$:
\begin{eqnarray}
& & f(r)=r+s{(r-R)}^k,\;\;s\in\Re,\;k\in N,\; k>2,\label{25}\\
& & m(r)=m+y{(r-R)}^b,\;\;y\in\Re,\;y>0,\;b\in N,\; b>2.\label{26}
\end{eqnarray}
As a preliminar consideration, note that for $k>2, b>2$ we have, for
$r\rightarrow R$, $P_r=o(E)$ and as a consequence condition $E+P_r\geq 0$ is always satisfied
near $r=R$ when $E\geq 0$.\\ 
With respect to (\ref{25})-(\ref{26}) and according to the reasonings above, 
for $r\rightarrow R$, three cases are possible:
\begin{enumerate}
\item $k=b$. We have $E\sim-2s(R-2m)(k^2-k){(r-R)}^k$ and for positivity of $E$ we must have 
$s<0$.\\ 
Moreover $E+P_{\phi}\sim -s(R-3m)k(k-1){(r-R)}^k$ is positive for $s<0$. 
Finally, $E+P_r+2P_{\phi}\sim k(k-1)(-Ry+ms){(r-R)}^k$ and positivity requires the unacceptable
condition $y<0$. Hence SEC is always violated near the matching surface $R$. 
\item $k>b$. In this case $E+P_{\phi}\sim -yRb(b-1){(r-R)}^b$ and with $y>0$ WEC is violated. Also 
SEC is violated with $y>0$. 
\item $b>k$. We have $E\sim -2sk(k-1)(R-2m){(r-R)}^k$ with the condition $s<0$ and
$E+P_{\phi}\sim -sk(k-1)(R-3m){(r-R)}^k$ is also positive for $s<0$. 
Finally, $E+P_r+2P_{\phi}\sim 2smk(k-1){(r-R)}^k$ and with $s<0$ SEC is violated.
\end{enumerate}
From reasonings above it follows that WEC, exception made for 
the case $2$, can be fulfilled near $r=R$. Conversely SEC is always violated. Hence,
matter content near $r=R$ has features similar to the the ones of dark energy,\\
The study above can be further generalised. In fact, the conclusions are the same with 
(\ref{25})-(\ref{26}) but with $\{k,b\}\in\Re$ with again $\{k,b\}>2$. Hence, we can arrive to
the same conclusions outlined above with (\ref{16})-(\ref{17}) substituted by:
\begin{eqnarray}
& & f_{\{n,b\}}(r)=r+{(r-R)}^k\sum_{h=0}^{n}\frac{f^{(h)}(R)}{h!}{(r-R)}^h,\nonumber\\
& & m_{\{n,b\}}(r)=m+{(r-R)}^b\sum_{h=0}^{n}\frac{m^{(h)}(R)}{h!}{(r-R)}^h,\label{27}\\
& & f(r)=f_{\{n,b\}}(r)+o({(r-R)}^{n+k}),\nonumber\\
& & m(r)=m_{\{n,b\}}(r)+o({(r-R)}^{n+b}).\label{28} 
\end{eqnarray}
Concerning energy conditions in the range
$r\in[R,\infty)$, the relevant equation in the slowly rotating limit for 
WEC is the following: 
\begin{equation}
E+P_r=-\frac{2}{f^2(r)}f_{,r,r}(r)\left[f(r)-2m(r)\right]\geq 0.
\label{29}
\end{equation} 
Since for regularity \cite{55} $f(r)>2m(r)$ , we must have $f_{,r,r}\leq 0$, i.e. $f(r)$ must be a non-convex
function for $r\in[R,\infty)$. Thanks to matching conditions (\ref{3}) we have 
$f_{,r}(R)=1$ and non-convex differentiable functions all necessarily have a non increasing 
$f_{,r}$. Hence, in the case $f(r)\neq r$, no asymptotically flat solutions, where
$\lim_{r\rightarrow\infty} f_{,r}=1$, are allowed. As a consequence, 
in the slowly rotating limit, the case with 
$f(r)\neq r$, in order to satisfy WEC, cannot be supported by an asymptotically flat isolated object.
Stated in other words, in the static and slowly rotating case,
a solution with a central Kerr black hole surrounded by an anisotropic field with 
a non-vanishing energy flow around the symmetry axis cannot be asymptotically flat.

\subsection*{General case}

In the full rotating case, computations are very cumbersome with respect to the slowly rotating limit.
In fact, for the energy conditions, $E$ must be replaced by $-\lambda_t$ given by (\ref{12}) in 
(\ref{9})-(\ref{11}). 
Nevertheless, in this section we show that sufficient conditions for WEC and SEC can
be done in a similar manner with respect to slowly rotating case. \\
As a first consideration, we have that 
$E(R)=-\frac{4BR(R^2-2Rm+a^2)}{{(R^2+a^2 x^2)}^2}$. Hence, also in the full rotating case 
condition $E(R)=0$ requires that $B=0$. Moreover, we have
$-\lambda_t(R)+P_{x}(R)=-2GR/{(R^2+a^2x^2)}$ and since $G\geq 0$, consequently we must take
$G=0$.\\
Concerning positive energy density condition, i.e. $E\geq 0$, 
with respect
to (\ref{18})-(\ref{19}), we have
\begin{equation}
E=-\frac{12RC}{(R^2+a^2 x^2)}\left[R^2-2mR+a^2\right]\left(r-R\right)+o(1),
\label{30}
\end{equation}
with obviously $C\leq 0$. Moreover for $C=H=0$ we have that 
$E=-\frac{6RD}{{(R^2+a^2 x^2)}}{(r-R)}^2+o(1)$ with obviously $D<0$ with a phenomenology for
$E$ similar to the one for slowly rotating case at higher orders.\\
Concerning WEC and SEC, we can simplify calculations by noticing that expression for 
$-\lambda_t$ in (\ref{12}) contains a positive square root. Hence, for models
with real eigenvalues, we can study sufficient but not
necessary conditions for WEC and SEC (\ref{9})-(\ref{10}). For 
$-\lambda_t\geq 0$ with (\ref{18})-(\ref{19}) and $B=G=0$, we have condition
$E-P_{\phi}\geq 0$ with:
\begin{eqnarray}
& & E-P_{\phi}=\label{31}\\
& & =\frac{-6\left[-RH(R^2+a^2 x^2)+C\left(3R^3-5mR^2+a^2(2R-x^2(R-m))\right)\right]}{R^2+a^2 x^2}
(r-R)+o(1). \nonumber
\end{eqnarray}
From (\ref{31}), with $R>2m$  we deduce that $H\geq 0$ and $C\leq 0$. We have 
$-\lambda_t+{\lambda}_{\phi}\geq 0$ for models with real eigenvalues.
Instead of $-\lambda_t+P_x\geq0$ we have $E-P_{\phi}+2P_x$ with 
\begin{eqnarray}
& & E-P_{\phi}+2P_x=\label{32}\\
& & =\frac{6\left[-RH(R^2+a^2 x^2)-C\left(R^3-3mR^2+a^2(2R-x^2(R-m))\right)\right]}{R^2+a^2 x^2}
(r-R)+o(1).\nonumber
\end{eqnarray}
From (\ref{32}) with $R>4m, C<0, H>0$ we deduce, as the slowly rotating case, that by choosing models with the ratio $|C|/H>(R^2+a^2)/(R^2-3m+a^2(R+m))$ condition $E-P_{\phi}+2P_x\geq 0$ 
is certainly satisfied near $r=R$.\\
Finally, instead of $-\lambda_t+P_r$ we have
$E-P_{\phi}+2P_r$. Since $P_r=o(E-P_{\phi})$ for $r\rightarrow R$, the same conclusions of
(\ref{31}) follow.\\
Concerning SEC, instead of $-\lambda_t+\lambda_{\phi}+P_r+P_x\geq 0$ we should consider
$P_x+P_r\geq 0$, but since we expect violation of SEC as the slowly rotating case, we must use another technique. To this purpose, note that for $x=\pm 1$, i.e. at $\rho=0, z=\pm r$, we have 
$K=0$ and $-\lambda_t$ reduces to $E$. Hence, we can study for $x=\pm 1$ the inequality
$E+P_{\phi}+P_r+P_x\geq 0$ together with (\ref{18})-(\ref{19}):
\begin{equation}
(E+P_{\phi}+P_r+P_x)(r,x=\pm 1)=
-\frac{12\left[RH(R^2+a^2)-mC(R^2-a^2)\right]}{{(R^2+a^2)}^2}.
\label{33}
\end{equation}
From (\ref{33}), violation of SEC for $R>4m, C<0, H>0$ follows.
As a result, we deduce that for WEC and SEC near matching surface $R$,
results of slowly rotating case still hold.\\
As a final step we must consider equation (\ref{29}). The inequality to study is 
$-\lambda_t+P_r\geq 0$ in the full range $r\geq R$. Since expression involved is rather cumbersome, it is not a simple task. However we can perform an argument similar to (\ref{33}) to show that rotation can amend the fate of the inequality $E+P_r\geq 0$ for asymptotically flat spacetimes in the slowly rotating limit. To this purpose we have that $-\lambda_t+P_r=E+P_r$ for $x=\pm 1$ with:
\begin{eqnarray}
& & E(r,x=\pm 1)+P_r(r,x=\pm 1)=\label{34}\\
& & =-\frac{2}{{(f^2+a^2)}^3}\left[f^2-2mf+a^2\right]
\left[ff_{,r,r}(a^2+f^2)+a^2(f_{,r}^2-1)\right]\geq 0.\nonumber
\end{eqnarray} 
Matching conditions (\ref{3}) impose that $f_{,r}=1$. For $f$ convex with 
$f_{,r,r}>0$, $f_{,r}$ is increasing and thus (\ref{34}) is violated, at least in a suitable
neighbourhood of $R$. Consequently, near $r=R$ $f$ must be a non-convex function with $f_{,r}$
decreasing. If we impose to have an asymptotically flat solutions, we must have
$f_{,r}(r\rightarrow\infty)=1$. Without the term involving $f_{,r}^2$ in (\ref{33}) we have no possibility since for a non-convex differentiable function $f_{,r}$ is decreasiang and cannot reach
the value $1$ at spatial infinity. However, with (\ref{33}), it is possible to build solutions with $f$
becoming convex at a given point $r_*>R$, thus with $f_{,r}^2-1<0$ but with $f_{,r}$ increasing
and approaching $1$ from below at spatial infinity. For models with 
$f_{,r,r}f(a^2+f^2)+a^2(f_{,r}^2-1)<0$ for $r>r_*$, inequality (\ref{34}) can be fulfilled. The reasoning above show that rotation could cure the slowly rotating case thus allowing to build asymptotically flat solutions satisfying WEC. However, note that since in our universe galaxies are living in an expanding
non-flat spacetime, models of galaxies that are non-asymptotically flat are physically viable.\\
As a final consideration of this section, note that for models 
with $C\neq 0$, from (\ref{30}) we have $E_{,r}(R)>0$. More generally, condition 
$E(R,x)=0$ and the positivity energy-density conditions does imply that $E_{,r}>0$ at least in a 
neighbourhood of $R$. As a result, $E(r,x)$ is a monotonically increasing function with respect to 
$r$ near $R$ and a typical spike profile is expected for $E$, similarly to the profile 
present in \cite{42}. This fact will be further commented at the conclusions.

\section{Energy conditions: the case $f(r)=r$}

For the case with $f(r)=r$ relations (\ref{6})-(\ref{8}) reduce to \cite{10}:
\begin{eqnarray} 
& & E+P_r=0,\;\;P_{\phi}=P_x,\;\;K=0,\label{34a}\\
& & E=\frac{2r^2 m_{,r}}{{(r^2+a^2 x^2)}^2},\label{35}\\
& & P_{\phi}=
-\frac{\left[r m_{,r,r}(r^2+a^2 x^2)+2a^2x^2 m_{,r}\right]}
{{(r^2+a^2 x^2)}^2}.\label{36}
\end{eqnarray}
Positive energy density, thanks to (\ref{35}), requires that $m_{,r}\geq 0$. For WEC we must thus consider:
\begin{eqnarray}
& & E+P_{\phi}=\label{37}\\
& & =-\frac{\left[r m_{,r,r}(r^2+a^2 x^2)+(2a^2 x^2-2r^2)m_{,r}\right]}
{{(r^2+a^2 x^2)}^2}.\nonumber
\end{eqnarray}
Since $m_{,r}(R)=0$ for matching conditions (\ref{3}), in a suitable neighbourhood of $R$
we have $m_{,r}>0$, i.e. $m_{,r}$ is an increasing function near $R$. Consequently we must have
$m_{,r,r}>0$ (convex function) in such a neighbourhood of $R$ with $E+P_{\phi}<0$. WEC is thus violated 
at least near $r=R$. It should be noted that the same phenomenon does appear if we consider the so called regular black hole solutions (see for example \cite{44} and references therein) near
$r=0$. In this case, by setting
$m(r)=Ar^s+o(1), s\geq 2$, we see that condition $E+P_{\phi}\geq0$ is fulfilled only for 
$A<0$, i.e. for $m_{,r}<0$ that is obviously unacceptable.\\
The reasings above show that, in order to fulfill WEC with (\ref{34})-(\ref{36}) a further field must be 
introduced, as for example an electromagnetic field.   
 
\section{Building solutions with $f(r)=r$ satisfying DEC}

As stated at the end of section above, for spacetimes with $f(r)=r$, further components must be added to the energy momentum tensor in order to satisfy WEC. As an interesting example, we can consider the central black hole depicted in terms of Kerr-Newman with a non vanishing electric charge $Q$ and with
outer event horizon $R_H$ located at $R_H=m+\sqrt{m^2-a^2-Q^2}$.
Moreover, in order to descrive a dark matter halo near $R$, all the boundaries conditions for $m(r)$ imposed in this paper are still valid. We must only perform on the metric (\ref{1}) the transformation
$m(r)\rightarrow m(r)-Q^2/(2r)$.  Hence, instead of (\ref{35})-(\ref{37}) we have:
\begin{eqnarray} 
& & E=\frac{2r^2 m_{,r}+Q^2}{{(r^2+a^2 x^2)}^2},\label{38}\\
& & P_{\phi}=
-\frac{\left[r m_{,r,r}(r^2+a^2 x^2)+2a^2x^2 m_{,r}-Q^2\right]}
{{(r^2+a^2 x^2)}^2}.\label{39}\\
& & E+P_{\phi}=\label{40}\\
& & =-\frac{\left[r m_{,r,r}(r^2+a^2 x^2)+(2a^2 x^2-2r^2)m_{,r}-2Q^2\right]}
{{(r^2+a^2 x^2)}^2},\nonumber
\end{eqnarray}
with (\ref{34a}) left unchanged. From (\ref{39})-(\ref{40}) it is easy to see that the presence of $Q$, contrary to the case with $Q=0$, certainly allow to fulfill WEC and SEC 
at least near $r=R$. Concerning DEC, we have:
\begin{eqnarray}
& & \frac{P_{\phi}}{E}=\label{41}\\
& & =-\frac{r m_{,r,r}(r^2+a^2 x^2)+2a^2x^2 m_{,r}-Q^2}{2r^2 m_{,r}+Q^2}\nonumber
\end{eqnarray}
For resonings above is easy to see that DEC is at least satisfied near $R$. For (\ref{19})
with $G=0$ we have $P_{\phi}/E=1$ for $r\rightarrow R$. Since $E=-P_r$ this means that at
the matching surface $R=R_{ms}$ modes propagate at the speed of light. In the following example we show how it is possible to easily build models satisfying at least WEC and DEC. To this purpose, we consider the following model:
\begin{equation}
m(r)=m+M\frac{{(r-R)}^3}{{(r+M)}^3},\label{42}
\end{equation}
where m is the black hole mass and M is the total mass of the matter content surrounding the black hole up to spatial infinity. The solution (\ref{43}) is thus asymptotically flat. As a numerical example, we 
can adopt for $M$ the infered mass surrounding the SMBH of our galaxy, Sgr A*.
Hence, we can set $m=1$, $M\sim 10^6$ and $R\sim 4$. As a title of example we can also set $a=Q=1/2$. 
In Fig. 1
we plot $E$, in Fig. 2 WEC and in Fig. 3 DEC. As a result, WEC and DEC are satisfied for 
$r\geq 4$. Concerning SEC, i.e. $P_{\phi}\geq$,  we see that it is satisfied up to 
$r\simeq 480$ and for $r\geq r*\sim 10^6$ where $P_{\phi}$ is positive.\\  
Similar results have been obtained in \cite{48}-\cite{50} but in the context of regular rotating black 
holes asymptotically matching Kerr-Newman solution. Is is worth to note that practically all models attempting to mimics rotation curves of galaxies, as for example the ones present in \cite{34}, violate
WEC at least $r=0$. For models with $f(r)=r$ a further component, for example an electromagnetic 
field, must be added \cite{50}.
\begin{figure}
\centering
\includegraphics[scale=0.35]{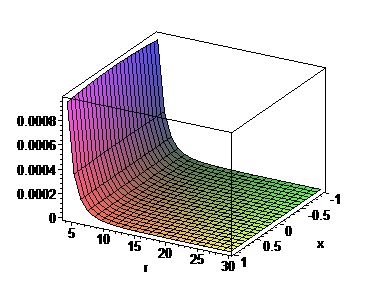} \quad
\caption{In figure plot of $E$ for $R=4, Q=1/2, a=1/2$.} 
\label{F1}
\includegraphics[scale=0.35]{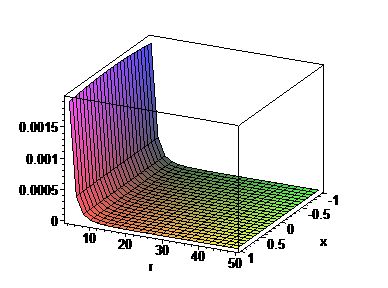}\\ 
\caption{In figure plot of $E+P_{\phi}$  for $R=4, Q=1/2, a=1/2$.}
\label{F2}
\includegraphics[scale=0.35]{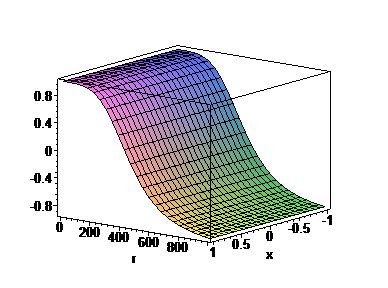} \quad
\caption{In figure plot of $P_{\phi}/E$  for $R=4, Q=1/2, a=1/2$.}
\label{F3}
\end{figure}

\section{Building solutions with $f(r)\neq r$ satisfying DEC}

In this section we consider models with $f(r)\neq r$. As argumented in the slowly rotating case, 
for asymptotically flat spacetimes WEC $E+P_r\geq 0$ cannot be fulfilled. The partial case
given by (\ref{34}) shows that in the rotating case there is hope to fulfill WEC. To this purpose, it
is necessary a non-convex expression for $f(r)$ up to some radius $r*$ and a convex one for $r>r*$. 
However, since any SMBH is embedded in an expanding non asymptotically flat uiverse, the non asymptotically flatness is not a physical severe problem.
As an example, we could manage $f(r)$ and $m(r)$ in order to have a spacetime equipped with a cosmological constant $\Lambda$. As an illustrative example we can take:
\begin{equation}
f(r)=r-s\frac{{(r-R)}^3}{r^2},\;\;r\geq R,\;\;s\in(0,1),
\label{43}
\end{equation}
together with (\ref{42}). 
Aymptotically for (\ref{1}) we have
\begin{equation}
ds^2=-dt^2+dr^2+{(1-s)}^2 r^2\left[d\theta^2+\sin^2(\phi)\;d \phi^2\right].
\label{44}
\end{equation}
Metric (\ref{44}) has an angle deficit (topological defect, see
\cite{56,57} and references therein) both with respect to $\theta$ and
$\phi$ with a positive string tension $1-\mu=1-s$. For such a model, we have verified that WEC and
DEC are generally satisfied in the full rotating case. In Fig. 4-11 we plotted WEC and DEC near $r=R$, since is the range where relativistic effects of SMBH are expected to be strong. Moreover, SEC is violated for
$r\in[4,r*], r*\sim 10^6$ and satisfied for $r>r*$.\\ 
As a final consideration, note that typical spike behavior of $E$ given in Fig 5. This behavior is similar to the one obtained in \cite{42} and is due to the physical requirement $E(R,x)=0$ and the positive 
energy density condition.  
\begin{figure}
\centering
\includegraphics[scale=0.35]{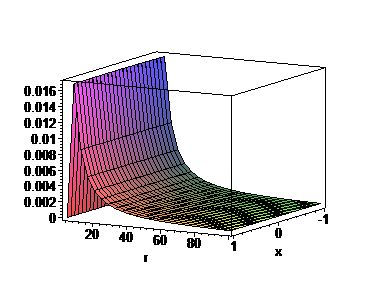} \quad
\caption{In figure the plot of ${(-\lambda_t)}$ for $R=4, a=1/2$.} 
\label{F4}
\includegraphics[scale=0.35]{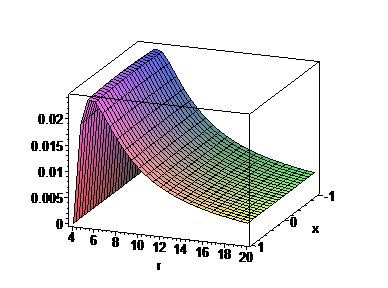}\\ 
\caption{In figure the plot of $E$ for $R=4, a=1/2$.}
\label{F5}
\includegraphics[scale=0.35]{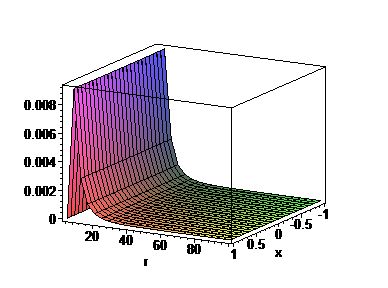}
\caption{In figure the plot of ${(-\lambda_t+P_r)}$ for $R=4, a=1/2$.}
\label{F6}
\includegraphics[scale=0.35]{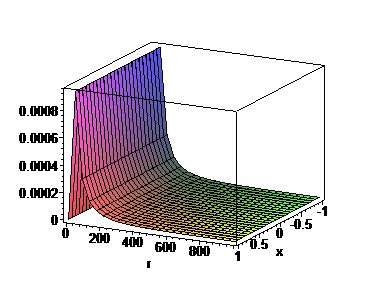}\\ 
\caption{In figure the plot ${(-\lambda_t+P_x)}$ for $R=4, a=1/2$.}
\label{F7}
\end{figure}
\begin{figure}
\centering
\includegraphics[scale=0.35]{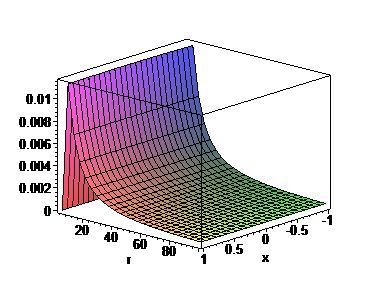} \quad
\caption{In figure the plot of $(-\lambda_t)+\lambda_{\phi}$ for $R=4, a=1/2$.} 
\label{F8}
\includegraphics[scale=0.35]{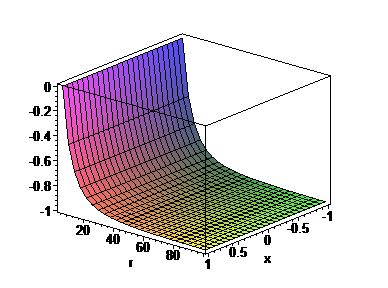}\\ 
\caption{In figure the plot of $P_{r}/(-\lambda_t)$ for $R=4, a=1/2$.}
\label{F9}
\includegraphics[scale=0.35]{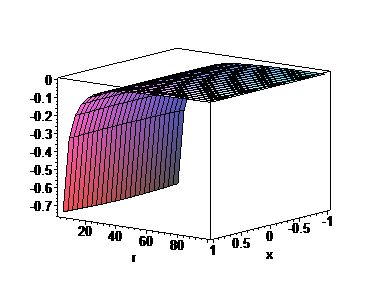}
\caption{In figure the plot of $\lambda_{\phi}/(-\lambda_t)$ for $R=4, a=1/2$.}
\label{F10}
\includegraphics[scale=0.35]{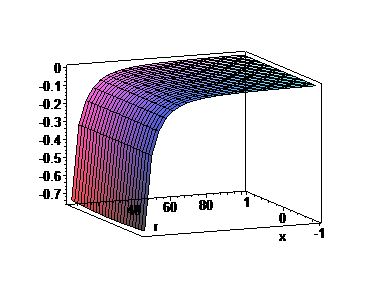}\\ 
\caption{In figure the plot $P_{x}/(-\lambda_t)$ for $R=4, a=1/2$.}
\label{F11}
\end{figure}

\section{Conclusions and final remarks}

Motivated by increasing interest in modeling galaxies with a central SMBH surrounded by a 
dark matter halo, we study the problem focusing in particular on energy conditions. The central SMBH
is often modeled by means of regular black hole solutions, as for example in \cite{34}. In fact, in order to mimic rotation curves for spiral galaxies, a Newtonian profile for $E$ is chosen. With the application of the algorithm in (\ref{10}), a rotating solution with 
(\ref{34a})-(\ref{36}) is obtained. As shown in this paper and in \cite{44,48,49,50}, such a class of solutions present a macroscopic (non Planckian magnitudo) violation of WEC at least near $r=R$. Since regular black holes are introduced in order to avoid classical singularity, the presence of phantom matter seems to be possible only at a quantum level. As a consequence, it seems natural to explore the possibility that central SMBH is depicted in terms of the  Kerr vacuum solution. The central singularity could be eliminated by means of quantum 
(non commutative) models at Planck lengths \cite{51,52,53}. A possible Planckian solution for singularity 
reasonably cannot have effects at macroscopic level outside the event horizon and 
thus classical Kerr solution can reasonably depict the central SMBH.\\
From reasonings above, it is clear the necessity to study models with a central Kerr black hole surrounded by anisotropic matter. Obviously a matching is necessary at a radius $R$ that is chosen to be 
closed to the radius $R_{ms}$ representing
the unstable spheroidal orbit for a marginally bound particle. To our purposes, we adopt the recent new class of anisotropic rotating fluids in \cite{55}. These solutions are smoothly matched at $R$ with Kerr solution. As a consequence of matching and boundary conditions imposed, we show that models with 
$f(r)\neq r$ can satisfy WEC near $R$ but in slowly rotating case WEC is certainly violated for asymptotically flat spacetimes at some place $r*>R$. We also have partially shown that huge rotation can amend this situation. The possible discovery of non vacuum asymptotically flat spacetimes with 
$f(r)\neq r$ satisfying WEC and smoothly matched to the Kerr vacuum solution can be matter of further investigations. As an illustrative example, in section 6 we obtained a solution with an angle deficit and satisfying WEC and DEC everywhere.\\
We have also studied the subcase with $f(r)=r$, showing that violation of WEC is unavoidable near
the matching surface $R$. To amend this situation and according to \cite{50} for 
regular black holes, a further field must be introduced as, for example, an electromagnetic one.
To this purpose we shown that the simple introduction of an electric charge $Q$ allows to obtain solutions satisfying at least WEC. Also in this subcase, we have presented a solution equipped with a non-vanishing 
charge $Q$, smoothly matched to the Kerr-Newman solution, and satisfying WEC and DEC.\\
As an interesting fact, it is worth to be noticed the difference between the plots of $E$ given by Fig. 1 and Fig. 5. In fact, for models with $E(R,x)=0$ the spike-like behavior of Fig. 5 is always expected near 
$R$. Conversely, for models with $E(R)\neq 0$ satisfying the positive energy density condition, 
the cusp-like behavior of Fig. 1 is expected. As a consequence, the shape of $E$ near $R$ can give 
informations on the nature of matter content near the SMBH of a given galaxy. The class of solutions given by (\ref{1}) can be an usefull tool to explore new possible galaxy models with a SMBH surronded by 
dark matter halo. This can be certainly matter for further investigations.

\end{document}